\documentclass[twocolumn,aps,prb,reprint,
superscriptaddress
]{revtex4-1}
\usepackage[pdftex]{graphicx}
\usepackage{amsmath,amssymb}
\usepackage[pdftex,
            colorlinks=true,
            citecolor=blue,
            linkcolor=blue]{hyperref}

\graphicspath{{./fig/}{./fig_phase_diagram_t20/}{./fig_phase_diagram_t202/}}

\usepackage{braket}
\usepackage{times,multirow,amsfonts,bm,xspace,pifont}
\usepackage{soul}
\usepackage[normalem]{ulem}
\usepackage{comment}

\begin{document}
\newcommand{\rr}{{\bm r}}
\newcommand{\q}{{\bm q}}
\renewcommand{\k}{{\bm k}}
\newcommand*\wien    {\textsc{wien}2k\xspace}
\newcommand*\textred[1]{\textcolor{red}{#1}}
\newcommand*\textblue[1]{\textcolor{blue}{#1}}
\newcommand{\ki}[1]{{\color{red}\st{#1}}}
\newcommand{\sgn}{\mathrm{sgn}\,}
\newcommand{\tr}{\mathrm{tr}\,}
\newcommand{\Tr}{\mathrm{Tr}\,}
\newcommand{\GL}{{\mathrm{GL}}}
\newcommand{\talpha}{{\tilde{\alpha}}}
\newcommand{\tbeta}{{\tilde{\beta}}}
\newcommand{\mathN}{{\mathcal{N}}}
\newcommand{\mathQ}{{\mathcal{Q}}}
\newcommand{\bv}{{\bar{v}}}
\newcommand{\bj}{{\bar{j}}}

\newcommand{\YY}[1]{\textcolor{magenta}{#1}}
\newcommand{\KN}[1]{\textcolor{green}{#1}}
\newcommand*{\KNS}[1]{\textcolor{green}{\sout{#1}}}
\newcommand*\YYS[1]{\textcolor{magenta}{\sout{#1}}}
\newcommand{\RH}[1]{\textcolor{blue}{#1}}
\newcommand*{\RHS}[1]{\textcolor{blue}{\sout{#1}}}
\newcommand{\reply}[1]{\textcolor{red}{#1}}
\newcommand{\replyS}[1]{\textcolor{red}{\sout{#1}}}

\title{Superconductivity in UTe$_2$ from local noncentrosymmetricity}

\author{Ryuji Hakuno}
\email[]{hakuno.ryuji.46v@st.kyoto-u.ac.jp}
\affiliation{Department of Physics, Graduate School of Science, Kyoto University, Kyoto 606-8502, Japan}
\author{Youichi Yanase}
\affiliation{Department of Physics, Graduate School of Science, Kyoto University, Kyoto 606-8502, Japan}
\date{\today}

\begin{abstract}
Superconductivity in UTe$_{2}$ has garnered significant attention, as it is widely recognized as a promising candidate for a spin-triplet superconductor. 
However, the symmetry of superconductivity and the microscopic origin of spin-triplet pairing remain subjects of debate. Nevertheless, various experiments imply an intimate coupling between magnetism and superconductivity. 
In this paper, we analyze a multi-sublattice periodic Anderson model that incorporates a spin-orbit coupling allowed in locally noncentrosymmetric crystals to discuss magnetic fluctuations and superconductivity in UTe$_2$. 
Due to the sublattice-dependent spin-orbit coupling, magnetic fluctuations become anisotropic, and the spin degeneracy of superconducting states is lifted. 
Our calculations reveal anisotropic antiferromagnetic fluctuations along the $b$- and $c$-axes, anisotropic ferromagnetic fluctuations along the $a$-axis, and their coexistence. These can be tuned by the $f$-electron's level. 
Superconductivity in the $A_u$ representation is predominant for a wide range of parameters, whereas the $B_{2u}$ representation is almost degenerate and can be stabilized. The direction of the $d$-vector changes as we increase the spin-orbit coupling.
We discuss the consistency between our results and several experiments.

\end{abstract}
\maketitle

\section{Introduction}

Superconductivity in a heavy fermion system UTe$_{2}$ was discovered in 2019~\cite{ranNearlyFerromagneticSpintriplet2019} and has attracted significant interest due to indications for spin-triplet superconductivity and the potential realization of topological superconductivity~\cite{Aoki_2022}. Experimental observations indicating spin-triplet superconductivity include extremely large upper critical magnetic fields that exceed the Pauli limit for all crystal axes~\cite{ranNearlyFerromagneticSpintriplet2019,Aoki_2019} and tiny reduction of the nuclear magnetic resonance (NMR) Knight shift in the superconducting state\cite{matsumura2023Large,ranNearlyFerromagneticSpintriplet2019}.
Another striking observation is the multiple superconducting phases under pressure\cite{braithwaiteMultiple2019,aokiMultiple2020} and magnetic fields\cite{ran2019extreme,Lewin_2023,kinjo2023Change,rosuel2023FieldInduced,tokiwa2025}, which indicate the presence of multicomponent order parameters of superconductivity, as in the representative spin-triplet superfluid $^{3}$He~\cite{Leggett_review} and superconductor UPt$_{3}$~\cite{Upt3_review}.
The evidence for spin-triplet superconductivity naturally implies that UTe$_{2}$ is a candidate for a topological superconductor\cite{ishizuka2019InsulatorMetal,teiPossible2023,yamazaki2025Higherorder}, which attracts attention as a platform for boundary Majorana fermions. 
However, the topological property strongly depends on the symmetry of superconductivity and the topology of Fermi surfaces\cite{ishizuka2019InsulatorMetal,yu2022Majorana,tei2025Topological}, which remain elusive despite many efforts to clarify them.

Although early-stage studies suggested breaking of the time-reversal symmetry in the superconducting state\cite{jiao2020Chirala,hayes2021Multicomponent}, recent experiments do not support this interpretation\cite{ajeesh2023Fate,Gu_science2025}. Consistent with the latter, several measurements rule out a double phase transition at ambient pressure and zero magnetic field\cite{Aoki_2022,rosa2022Single,theuss2024Singlecomponent}. On the other hand, magnetic penetration depth measurement has reported power-law scaling consistent with the time-reversal symmetry broken $B_{3u}+iA_u$ state\cite{ishiharaChiralSuperconductivityUTe22023}.

The symmetry of unconventional superconductivity is closely related to the mechanism of Cooper pairing. 
A canonical mechanism of spin-triplet superconductivity relies on ferromagnetic fluctuations~\cite{Leggett_review}, and the quantum critical behaviors in UTe$_2$ suggested the presence of ferromagnetic fluctuations\cite{ranNearlyFerromagneticSpintriplet2019}. Thus, it was considered that the pairing interaction in UTe$_2$ would be mediated by ferromagnetic fluctuations. 
However, subsequent neutron scattering experiments detected antiferromagnetic fluctuations with the wave vector $\bm{q} \simeq (0,\pi,0)$ rather than ferromagnetic fluctuations\cite{duan202incommensurate,knafo2021Lowdimensional}, and so far direct evidence for ferromagnetic fluctuations has not been obtained.
Since antiferromagnetic fluctuations usually enhance the spin-singlet superconductivity rather than the spin-triplet superconductivity~\cite{yanase2003Theory}, the need to reconsider the mechanism of superconductivity has been recognized.

To theoretically investigate the symmetry and mechanism of superconductivity, it is indispensable to elucidate the underlying electronic structure. It is also essential to study the role of the spin-orbit coupling (SOC) because spin-triplet superconducting states degenerate with respect to the spin degree of freedom when the SOC is absent. 
Since the discovery of superconductivity in UTe$_2$, numerous experiments and first-principles calculations have been carried out to elucidate the electronic structure of UTe$_{2}$.
In the first-principles band calculations~\cite{ishizuka2019InsulatorMetal,xu2019QuasiTwoDimensional,miao2020Low,shishidouTopological2021,harima2020How,shick2019Spinorbita,shick2021mathrma,choi2024Correlated,Halloran2025,shimizu2025Electronic,Aoki_2019}, an insulating band structure was obtained by the naive LDA calculation~\cite{Aoki_2019}, which is inconsistent with the experiments. However, the insulator-metal transition occurs due to a finite Coulomb interaction $U$ and the dimensionality of the Fermi surfaces varies depending on the magnitude of $U$, as shown by the DFT+$U$ calculation~\cite{ishizuka2019InsulatorMetal}. The two-dimensional Fermi surfaces are obtained for large $U$ as revealed also by the dynamical mean field theory combined with DFT calculations~\cite{miao2020Low,xu2019QuasiTwoDimensional}. 
In experimental studies, angle-resolved photoemission spectroscopy (ARPES)\cite{fujimori2019Electronic,miao2020Low,fujimori2021CoreLevel} has been used to study the band structure below the Fermi level. Two-dimensional Fermi surfaces and the signature of a three-dimensional one have been reported~\cite{miao2020Low}.
After the development of the method for synthesizing ultra-clean single crystals~\cite{Sakai2022crystal}, quantum oscillations\cite{aoki2022First,aoki2023Haas,aoki2024High,broyles2023Revealing,eaton2024Quasi2Da} and quantum interference\cite{weinberger2024Quantum} have been successfully observed, and it is becoming increasingly evident that the Fermi surface is quasi-two-dimensional.
However, we should pay attention to the effects of surfaces in the ARPES measurements and of high magnetic fields in quantum oscillation measurements when we discuss the electronic structure in bulk at zero magnetic field.
Rather isotropic electric resistivity\cite{eo2022axis}, magnetic penetration depth~\cite{ishiharaChiralSuperconductivityUTe22023}, and superconducting coherence length\cite{suetsugu2024Fully} suggest presence of a Fermi surface of three-dimensional nature, and it is also obtained in some calculations using the dynamical mean field theory~\cite{choi2024Correlated,Halloran2025}. 

Several theoretical studies have been conducted to elucidate the relationship between magnetism and superconductivity in UTe$_{2}$\cite{ishizuka2021Periodica,kreisel2022Spintriplet,hazra2023Tripleta,yu2023Theory,choi2024Correlated,haruna2024Possible,tei2024Pairing}. However, the central issues have not yet been fully resolved.
In some studies based on microscopic models, ferromagnetic fluctuations were derived, and it was demonstrated that they stabilize spin-triplet superconductivity~\cite{ishizuka2021Periodica,choi2024Correlated}.
There are also studies that connect spin-triplet superconductivity to antiferromagnetic fluctuations; however, these studies assume spin fluctuations and their coupling to electrons phenomenologically\cite{kreisel2022Spintriplet,tei2024Pairing}.
The purpose of this theoretical study is to get insight into the mechanism of superconductivity and to propose potential superconducting symmetry by analyzing an effective model for UTe$_2$. 
In our previous study, we constructed a mixed-dimensional periodic Anderson model with coexisting one- and three-dimensional electronic structures\cite{hakuno2024Magnetism}. This model shows spin-triplet superconductivity due to an effective interaction mediated by antiferromagnetic fluctuations that is consistent with the neutron scattering experiments~\cite{duan202incommensurate,knafo2021Lowdimensional}.
However, the SOC was neglected and, therefore, the spin degree of freedom of spin-triplet Cooper pairs, which is represented by the $d$-vector~\cite{Leggett_review}, was not discussed. To resolve this issue, we construct and analyze an extended model.

In this paper, we discuss the magnetic anisotropy and the $d$-vector of the superconducting order parameter by considering the SOC that inevitably arises from the crystal structure of UTe$_{2}$. There are two sublattices in the unit cell, and the U atoms form a ladder structure. Thus, the local inversion symmetry is broken at the U sites, and the antisymmetric SOC appears, as is expected in locally noncentrosymmetric compounds~\cite{LNCSC_review} such as CeRh$_2$As$_2$ and transition metal dichalcogenides. 
Taking into account the SOC, we construct and analyze the multi-sublattice periodic Anderson model.
In Sec.~II, we introduce the model and method. Next, in Sec.~III, we show the magnetic anisotropy of ferromagnetic and antiferromagnetic fluctuations. Then, superconductivity is discussed in Sec.~IV. Finally, in Sec.~V, the consequences of local inversion symmetry breaking in magnetism and superconductivity are summarized, and issues left for future studies are discussed.

\section{Model and Method}
We construct a $f$-$d$-$p$ periodic Anderson model with two sublattices and the staggered SOC.
In addition to the quasi-two-dimensional Fermi surfaces, whose existence has been verified\cite{aoki2023Haas,aoki2022First,eaton2024Quasi2Da,miao2020Low,broyles2023Revealing}, we assume a three-dimensional $f$-electron Fermi surface, as suggested by some experiments\cite{suetsugu2024Fully,miao2020Low,eo2022axis} and calculations\cite{ishizuka2019InsulatorMetal,shishidouTopological2021,shimizu2025Electronic,roising2024Thermodynamic,choi2024Correlated}.
The total Hamiltonian of the system is as follows.
\begin{align}
    H=H_0+H^{f}_\mathrm{ASOC}+H_\mathrm{int}.
\end{align}
The first term $H_0$ contains the kinetic energy and hybridization terms of the $f$, $d$, and $p$ electrons.
The matrix form of the Hamiltonian is as follows.
\begin{align}
H_0=
\begin{pmatrix}
\varepsilon^\mathrm{intra}_f & \varepsilon^\mathrm{inter}_f & V_{f-d} & 0 & V_{f-p} & -V_{f-p} \\
 & \varepsilon_f^\mathrm{intra} & 0 & V_{f-d} & -V_{f-p} & V_{f-p} \\
 &  & \varepsilon_d^\mathrm{intra} & \varepsilon^\mathrm{inter}_d & V_{d-p} & -V_{d-p} \\
 &  &  & \varepsilon_d^\mathrm{intra} & -V_{d-p} & V_{d-p} \\
 & h.c. &  &  & \varepsilon_p^\mathrm{intra} & \varepsilon^\mathrm{inter}_p \\
 &  &  &  &  & \varepsilon_p^\mathrm{intra} \\
\end{pmatrix}.
\end{align}
Here, we show the $6 \times 6$ matrix for the orbital and sublattice degrees of freedom and omit the unit matrix for the spin degree of freedom.
Although hybridization terms may have complicated momentum dependence, we assume momentum-independent hybridization parameters $V_{l-l'}$ ($l,l'=f,d,p$) for simplicity. 
In the intra-orbital part, $\varepsilon^\mathrm{intra}_l$ and $\varepsilon^\mathrm{inter}_l$ represent intra- and inter-sublattice hopping and are written as
\begin{align}
\varepsilon^\mathrm{intra}_f({\bm k}) =&-2t_x^f\cos k_x-2t_y^f\cos k_y-2t_z^f (\cos k_z +1) \notag \\
&+\Delta_f-\mu,\\
\varepsilon^\mathrm{intra}_d({\bm k}) =&-2t_x^d\cos k_x-2t_y^d\cos k_y+\Delta_d-\mu,\\
\varepsilon^\mathrm{intra}_p({\bm k}) =&2t_x^p\cos k_x+2t_y^p\cos k_y+\Delta_p-\mu,\\
\varepsilon^\mathrm{inter}_f({\bm k}) =&t_\mathrm{AB}^f+{t'}_\mathrm{AB}^{f}e^{ik_z},\\
\varepsilon^\mathrm{inter}_d({\bm k}) =&t_\mathrm{AB}^d+{t'}_\mathrm{AB}^{d}e^{ik_z},\\
\varepsilon^\mathrm{inter}_p({\bm k}) =&t_\mathrm{CD}^p+{t'}_\mathrm{CD}^{p}e^{ik_y}.
\end{align}
The second term $H^{f}_\mathrm{ASOC}$ shows a sublattice-dependent SOC\cite{maruyamaLocally2012,ishizuka2018Oddparity,LNCSCreview} of the $f$ electrons and can be written as
\begin{align}
    H^{f}_\mathrm{ASOC}=\sum_{\bm{k}ss'mm'}\bm{g}(\bm{k})\cdot\bm{\sigma}_{ss'}\otimes\tau_{mm'}^z c^{\dagger}_{f\bm{k}ms}
    c_{f\bm{k}m's'},
\end{align}
where $\sigma^\mu$ and $\tau^\mu$ represent the Pauli matrix for spin and sublattice, respectively.
In our model, this SOC term is the origin of anisotropic magnetic fluctuations and the lifting of spin degeneracy in spin-triplet superconducting states.
The sign of the SOC is opposite between the two sublattices because the system preserves the global inversion symmetry~\cite{LNCSCreview}.
Since the mirror symmetry in the $c$-axis direction is locally broken at the Uranium sites of UTe$_2$, the $\bm{g}$-vector takes the form,
\begin{align}
    \bm{g}(\bm{k})=(\alpha_1 \sin{k_y},\alpha_2 \sin{k_x},0),
\end{align}
where $\alpha_1$ and $\alpha_2$ are independent coupling constants. Because the 3-, 4-, and 6-fold rotation symmetries are absent, there is no symmetry constraint on the coupling constants. 
In the last term $H_\mathrm{int}$, we take into account the onsite Coulomb interaction of $f$ electrons, 
\begin{align}
    H_\mathrm{int}=U\sum_{i}n^{f}_{i\uparrow}n^{f}_{i\downarrow}.
\end{align}

As shown in the previous study~\cite{hakuno2024Magnetism}, an important parameter of the model is the energy level of the $f$ electrons $\Delta_f$. In addition, the coupling constants of the SOC $\alpha_1$, $\alpha_2$ are essential for the anisotropy of magnetic fluctuations and superconductivity.
We fix the other parameters as
$(t_x^f,t_y^f,t_z^f,t_x^d,t_y^d,t_x^p,t_y^p)=(0.08,0.035,0.06,1.2,0,0,1)$, $(t_\mathrm{AB}^f,{t'}_\mathrm{AB}^f,t_\mathrm{AB}^d, {t'}_\mathrm{AB}^d,t_\mathrm{CD}^p,{t'}_\mathrm{CD}^p)=(0.2,-0.1,1.5,-0.6,3,2)$,  
$(V_{f-d},V_{f-p},V_{d-p})=(0.1,0.1,0.5)$, and $(\Delta_d,\Delta_p,\mu)=(2.5,-3.3,-0.1)$, for which the non-interacting part of the model has the Fermi surfaces shown in Fig.~\ref{fig:gap_band}, which is similar to the DFT+$U$ calculations~\cite{ishizuka2019InsulatorMetal,shishidouTopological2021,shimizu2025Electronic} and the reduced models~\cite{hakuno2024Magnetism,liu2024densityfunctionaltheorybased}.

We employ the random phase approximation (RPA) and Eliashberg theory to calculate the correlation-enhanced spin susceptibility and superconducting instability, as used in numerous studies for strongly correlated electron systems\cite{yanase2003Theory}.
The non-interacting Green function is defined as follows:
\begin{align}
    \hat{G}(k)=\left(i \omega_n\hat{\bm{1}}_{12\times12} -\hat{H}_0(\bm{k}) -\hat{H}_{\rm ASOC}^{f}(\bm{k}) \right)^{-1},
\end{align}
where $\hat{\bm{1}}$ denotes the unit matrix and $\hat{H}_0(\bm{k})$ and $\hat{H}_{\rm ASOC}^{f}(\bm{k})$ are matrix representations of $H_0$ and $H_{\rm ASOC}^{f}$, respectively.
For simplicity, we write the three-dimensional wave vector and the Matsubara frequency $\omega_n=(2n+1)\pi T$ as $k=(\bm{k},i\omega_n)$.

We calculate the susceptibility of $f$ electrons, which is important for magnetism and superconductivity.
Hereafter, the intra-orbital $f$-electron component of the Green function is denoted by $G$, which contains the subscript for sublattice and spin.
The irreducible susceptibility is calculated as follows:
\begin{align}
    \chi_{s_1 s_2 s_3 s_4}^{(0)\; m_1 m_2}(q)=-\frac{T}{N}\sum_{k}
    G^{m_1 m_2}_{s_1 s_3}(k)G^{m_2 m_1}_{s_4 s_2}(k+q),
\end{align}
where $m_i=\mathrm{A,B}\;(i=1,2)$ is the index of sublattices for the Uranium sites, and $s_i=\uparrow,\downarrow\;(i=1,2,3,4)$ is the spin index.

The spin and charge susceptibility is derived from the generalized susceptibility and is obtained as follows:
\begin{align}
    \hat{\chi}(q)=(\hat{\bm{1}}_{8\times8}-\hat{\chi}^{0}(q)\hat{\Gamma}^{(0)})^{-1}\hat{\chi}^{0}(q),
\end{align}
with $\Gamma^0$ being the bare vertex of the onsite Coulomb interaction
\begin{equation}
  \Gamma^{(0)\; m_1 m_2}_{s_1 s_2 s_3 s_4}=
  \begin{cases}
    U & s_1=s_3\ne s_2=s_4, m_1=m_2 \\
    -U & s_1=s_2\ne s_3=s_4, m_1=m_2 \\
    0       & \text{otherwise}
  \end{cases}.
\end{equation}

We solve the linearized Eliashberg equation to study the superconducting instability. The linearized Eliashberg equation is given in the form of an eigenvalue equation,
\begin{align}
    \lambda\Delta^{m_1 m_2}_{s_1 s_2}(\bm{k})=\frac{T}{N}
    \sum_{\bm{k'}}V^{m_1 m_2}_{s_1 s_3 s_4 s_2}(\bm{k}-\bm{k}',i\Omega_m=0)
    F^{m_1 m_2}_{s_3 s_4}(\bm{k}').
\end{align}
In the RPA, the effective interaction $V$ is assumed to be mediated by spin, charge, and multipole fluctuations, and it is obtained as 
\begin{align}
    \hat{V}(q)=\frac{\hat{\Gamma}^{(0)}}{2}+\hat{\Gamma}^{(0)}\hat{\chi}(q)\hat{\Gamma}^{(0)}.
\end{align}
For simplicity, we ignore the Matsubara frequency dependence of the gap function $\Delta^{m_1 m_2}_{s_1 s_2}(\bm{k})$. The superconducting transition temperature is overestimated by this simplification, but the superconducting states are qualitatively unchanged~\cite{yanase2003Theory}.
Then, linearizing the anomalous Green function leads to 
\begin{align}
    \hat{F}(\bm{k})=\sum_{\omega_n}\hat{G}(\bm{k},i\omega_n)\hat{\Delta}(\bm{k})\hat{G}^{\top}(-\bm{k},-i\omega_n).
\end{align}

We classify the gap function into the eight irreducible representations (IR) of the $D_\mathrm{2h}$ point group and numerically calculate the superconducting eigenvalues $\lambda$.
The odd-parity IR $A_u$, $B_{1u}$, $B_{2u}$, and $B_{3u}$ correspond to the odd-parity spin-triplet superconductivity and the even-parity IR $A_g$, $B_{1g}$, $B_{2g}$, and $B_{3g}$ represent the even-parity spin-singlet superconductivity~\cite{Sigrist-Ueda}. 
Note that the definition of odd-parity IR $A_u$, $B_{1u}$, $B_{2u}$, and $B_{3u}$ is different from that in our previous study\cite{hakuno2024Magnetism}. This is because the classification scheme takes into account the spin degrees of freedom in this paper, while it was neglected in the previous study. The former is the standard classification scheme for superconductors with SOC. 
In the following results, we fix the Stoner factor, which is the largest eigenvalue of the susceptibility matrix, as $U\mathrm{max}_{q}\chi(q)=0.9$.
We take $64\times64\times64$ ${\bm k}$-meshes in the three-dimensional Brillouin zone.
The temperature is fixed as $T=0.001$.
For the Fourier transform of imaginary time and Matsubara frequency, we use the sparseIR package\cite{wallerberger2023sparseir}.

\section{Magnetic fluctuations}

\begin{figure*}[htbp]
    \centering
    \includegraphics[width=0.80\textwidth]{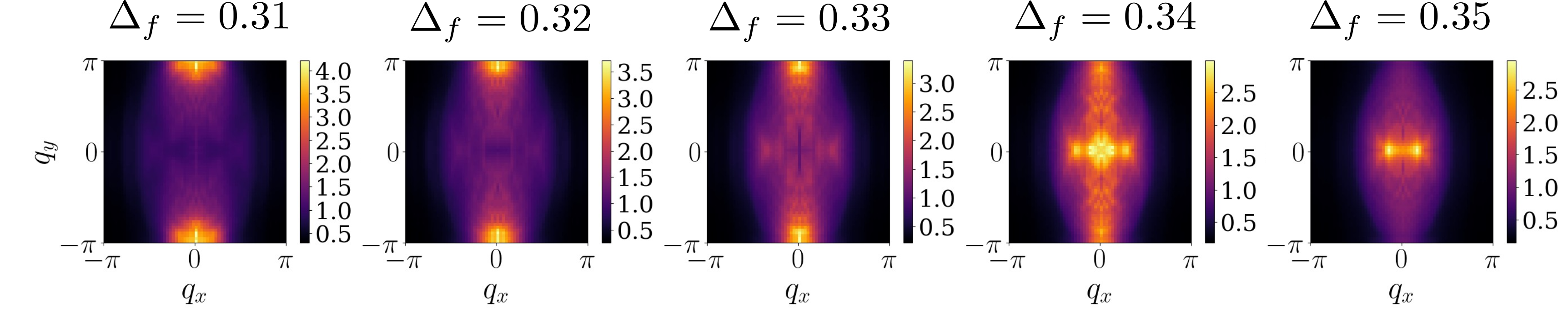}
    \caption{
    Momentum dependence of intra-sublattice spin susceptibility $\chi_\mathrm{AA}({\bm q}) = \chi_\mathrm{BB}({\bm q})$ on the $q_z=0$ plane without the SOC, $\alpha_1=\alpha_2=0$. The $f$-electron level $\Delta_f$ is shown on top of each figure. The antiferromagnetic fluctuation changes to the ferromagnetic one as increasing $\Delta_f$. 
    }
    \label{fig:Delta_f_dep}
\end{figure*}

 \begin{figure*}[htbp]
    \centering
    \includegraphics[width=0.90\textwidth]{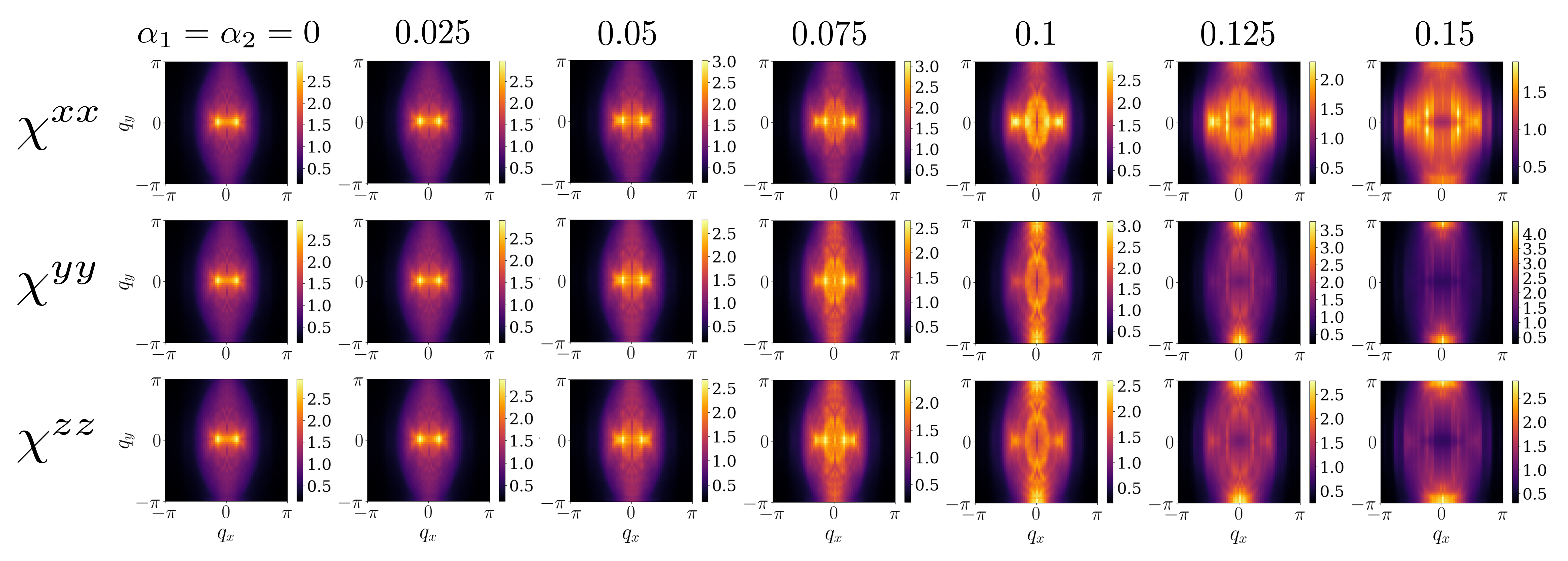}
    \caption{
    Spin susceptibility $\chi^{\mu\mu}_\mathrm{AA}({\bm q})$ along each crystalline axis $\mu=x,y,z$ with the SOC. We assume $\alpha_1 = \alpha_2$ and show their magnitude on top of each figure. We set $\Delta_f=0.35$ and present the momentum dependence on the $q_z=0$ plane.
    }
    \label{fig:susc_ferro}
\end{figure*}

\begin{figure*}[ht]
    \centering
    \includegraphics[width=0.90\textwidth]{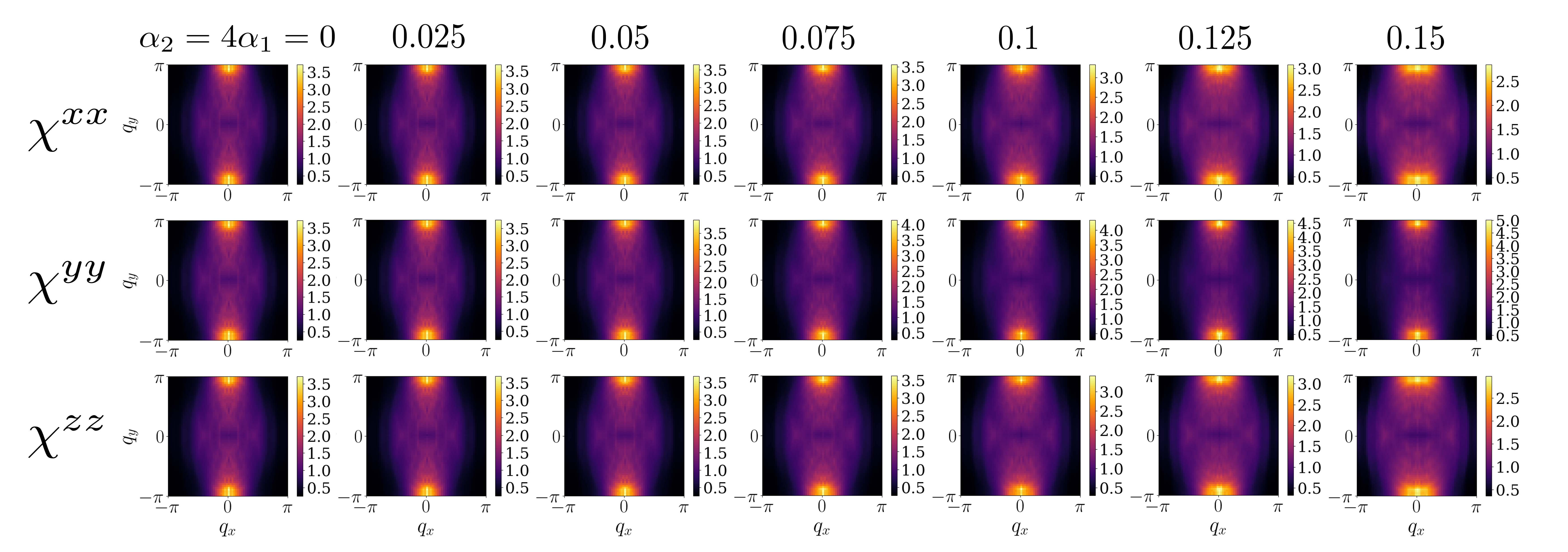}
    \caption{
    Spin susceptibility $\chi^{\mu\mu}_\mathrm{AA}({\bm q})$ with the SOC for $\Delta_f=0.32$. We assume $\alpha_2 = 4\alpha_1$ and show their magnitudes on top of each figure. 
    The momentum dependence on the $q_z=0$ plane is plotted.}
\label{fig:susc_antiferro}
\end{figure*}

In this section, we discuss the results of magnetic fluctuations.
The static spin susceptibility tensor is calculated as
\begin{align}
    \chi_{mm'}^{\mu\nu}(\bm{q})=\sigma_{s_1 s_2}^{\mu}
    \chi^{mm'}_{s_2 s_1 s_3 s_4}(\bm{q},i\Omega_m=0)\sigma_{s_3 s_4}^{\nu},
\end{align}
for the crystallographic directions $\mu,\nu=x,y,z \, (=a,b,c)$.
In the absence of the SOC, the magnetic correlation is isotropic in spin space: The diagonal part of the spin susceptibility tensor is the same for all directions $\chi^{xx}=\chi^{yy}=\chi^{zz}$, 
and the off-diagonal part vanishes.
We first show the results in our model without the SOC and later discuss the effects of the SOC. 

Magnetic fluctuations are significantly affected by the $f$ electron level $\Delta_f$, consistent with our previous study on the single-sublattice model~\cite{hakuno2024Magnetism}. We show the intra-sublattice spin susceptibility $\chi_\mathrm{AA}({\bm q})$ for various values of $\Delta_f$ in Fig.~\ref{fig:Delta_f_dep}.
We see the antiferromagnetic fluctuation with the wave vector $\bm{q} \simeq (0,\pi,0)$ in the small $\Delta_f$ region, which is consistent with the magnetic excitation observed by the neutron scattering experiments~\cite{duan202incommensurate,knafo2021Lowdimensional}. However, the ferromagnetic fluctuation is dominant in the large $\Delta_f$ region.
These behaviors are consistent with our previous study\cite{hakuno2024Magnetism}.
The antiferromagnetic fluctuation originates from the Fermi surface nesting of the $f$-electron bands, and the ferromagnetic fluctuation is probably related to the three-dimensional property of the Fermi surface.
The inter-sublattice spin susceptibility $\chi_\mathrm{AB}({\bm q})=\chi_\mathrm{BA}({\bm q})$ is small compared to the intra-sublattice one, and shows the qualitatively same momentum dependence.
The inter-sublattice spin susceptibility is positive, indicating that the inter-ladder magnetic correlation is ferromagnetic in both small and large $\Delta_f$ regions.

Next, we investigate the roles of the SOC.
We perform calculations by varying the SOC strength with fixing the $f$ electron level $\Delta_f$. The typical cases of dominant antiferromagnetic or ferromagnetic fluctuations are discussed in the following.
First, Fig.~\ref{fig:susc_ferro} shows the diagonal spin susceptibility for $\Delta_f = 0.35$, 
for which the ferromagnetic fluctuation is dominant in the absence of the SOC.
When the SOC is small $\alpha_1=\alpha_2\leqq0.05$, we see peaks of spin susceptibilities at $\bm{q}=0$ for all directions and Ising anisotropy with the easy axis along the crystallographic $a$ direction. This magnetic anisotropy is caused by the SOC term proportional to $\alpha_1$ and is not changed by the other $\alpha_2$ term unless $\alpha_1 \ll \alpha_2$.
The magnetic anisotropy with easy $a$-axis is consistent with experimental results that have reported the Ising ferromagnetic fluctuation\cite{ranNearlyFerromagneticSpintriplet2019,sundar2019Coexistencea,tokunaga2019125TeNMRa,ambika2022Possible,fujibayashi2023LowTemperaturea}, but the antiferromagnetic fluctuation observed in neutron scattering experiments is not reproduced in this case.
For large SOC $\alpha_1=\alpha_2>0.05$, the maximum position of spin susceptibility along the $b$ and $c$ axes changes from $\bm{q}=\bm{0}$ to $\bm{q}=(0,\pi,0)$, while the magnetic fluctuation along the $a$ axis remains almost ferromagnetic.  
Thus, the Ising ferromagnetic fluctuation along the $a$ direction and the antiferromagnetic fluctuations along the other directions coexist in this case. The former is consistent with the magnetic anisotropy of uniform susceptibility~\cite{ranNearlyFerromagneticSpintriplet2019,Aoki_2022}, and the latter could be consistent with the neutron scattering experiments~\cite{duan202incommensurate,knafo2021Lowdimensional}. 

Next, we show the results for $\Delta_f=0.32$ for which the magnetic fluctuation is antiferromagnetic in the absence of the SOC. The SOC does not qualitatively change the momentum dependence of spin susceptibility for $\alpha_2=4\alpha_1$, but induces anisotropy in antiferromagnetic fluctuations as shown in Fig.~\ref{fig:susc_antiferro}.
The orientation of the magnetic easy axis is along the $b$ axis.
The antiferromagnetic spin susceptibility at ${\bm q} = (0,\pi,0)$ is the second largest in the $c$ axis, while it is the smallest in the $a$ axis. Thus, the magnetic anisotropy is different between ferromagnetic and antiferromagnetic fluctuations. This is 
not unusual in models of itinerant magnetism.
Although the magnetic field is set to zero in this calculation, a recent NMR study under the $b$-axis magnetic field has reported magnetic fluctuations in the $b$-axis direction\cite{tokunaga2023Longitudinal} and could be consistent with the above results.
For other ratios of $\alpha_1$ and $\alpha_2$, the wave vector of antiferromagnetic fluctuations can deviate from ${\bm q} = (0,\pi,0)$. For example, when $\alpha_1=\alpha_2$, we find a magnetic fluctuation around ${\bm q} = (\delta,\pi -\delta',0)$ ($\delta, \delta' \ll \pi$) and symmetric points. This wave vector is similar to the magnetic wave vector of the antiferromagnetic phase under high pressure~\cite{Knafo2025}. In this paper, we focus on UTe$_2$ at ambient pressure and do not consider this case in the following. 
 
 \section{Superconductivity}

\begin{table*}[htbp]
    \centering
    \caption{Basis functions of the superconducting gap function, $\psi^\mu$ and $\bm{d}^\mu$, for each IR in the point group $D_{2h}$.}
    \label{tab:Irreps}
    \begin{tabular}{cccc}
        \hline
        IR & $\tau_0,\tau_x$ & $\tau_y$ & $\tau_z$ \\ \hline
        $A_{1g}$ & $\psi({\bm k})=1$ & $\psi({\bm k})=k_z$ & $\bm{d}({\bm k})=c_1 k_y\hat{x}+c_2 k_x\hat{y}$ \\
        $B_{1g}$ & $\psi({\bm k})=k_x k_y$ & $\psi({\bm k})=k_x k_y k_z$ & $\bm{d}({\bm k})=c_1 k_x\hat{x}+c_2 k_y\hat{y}+c_3 k_z\hat{z}$ \\
        $B_{2g}$ & $\psi({\bm k})=k_x k_z$ & $\psi({\bm k})=k_x$ & $\bm{d}({\bm k})=c_1 k_z\hat{y}+c_2 k_y\hat{z}$ \\
        $B_{3g}$ & $\psi({\bm k})=k_y k_z$ & $\psi({\bm k})=k_y$ & $\bm{d}({\bm k})=c_1 k_z\hat{x}+c_2 k_x\hat{z}$ \\ 
        $A_{1u}$ & $\bm{d}({\bm k})=c_1 k_x\hat{x}+c_2 k_y\hat{y}+c_3 k_z\hat{z}$ & $\bm{d}({\bm k})=\hat{z}$ & $\psi({\bm k})=k_x k_y$ \\
        $B_{1u}$ & $\bm{d}({\bm k})=c_1 k_y\hat{x}+c_2 k_x\hat{y}$ & $\bm{d}({\bm k})=c_1 k_y k_z\hat{x}+c_2 k_x k_z\hat{y}+c_3 k_x k_y\hat{z}$ & $\psi({\bm k})=1$ \\
        $B_{2u}$ & $\bm{d}({\bm k})=c_1 k_z\hat{x}+c_2 k_x\hat{z}$ & $\bm{d}({\bm k})=\hat{x}$ & $\psi({\bm k})=k_y k_z$ \\
        $B_{3u}$ & $\bm{d}({\bm k})=c_1 k_z\hat{y}+c_2 k_y\hat{z}$ & $\bm{d}({\bm k})=\hat{y}$ & $\psi({\bm k})=k_x k_z$ \\\hline
    \end{tabular}
\end{table*}

In this section, we study the relationship between superconductivity and magnetic fluctuations.
We calculate the superconducting gap function by solving the linearized Eliashberg equation and decompose it as
 \begin{align}
     \Delta_{ss'}^{mm'}({\bm k})=\sum_{\mu}[(\psi^{\mu}({\bm k})+\bm{d}^{\mu}({\bm k})\cdot\bm{\sigma})i\sigma^y]_{ss'}\otimes\tau^{\mu}_{mm'},
 \end{align}
where $\psi^{\mu}(\bm{k})$ and $\bm{d}^{\mu}(\bm{k})$ are spin-singlet and spin-triplet pairing components, respectively.
The Pauli matrix in sublattice space $\tau^0$ and $\tau^z$ represent the intra-sublattice pairing components, while $\tau^x$ and $\tau^y$ are the inter-sublattice pairing components.
We normalize the gap function as $\sum_{\bm{k}\mu}(|\psi^{\mu}(\bm{k})|^2+|\bm{d}^{\mu}(\bm{k})|^2)=1$.
The superconducting states are classified by the point group symmetry.
The basis functions for the IRs in the point group $D_{2h}$ are described in Table~\ref{tab:Irreps}.

The inter-sublattice hopping term of $f$ electrons can be approximated by $t^f_\mathrm{AB}\tau^x$ because $t^f_\mathrm{AB}>|t'^f_\mathrm{AB}|$.
If we take a basis that diagonalizes this term, the $\tau^0$ and $\tau^x$ components of the superconducting gap function appear in the diagonal part representing the intra-band paring, while the $\tau^y$ and $\tau^z$ components correspond to the inter-band paring.
Since the inter-band pairing is negligible in the weak-coupling limit where the superconducting gap is infinitesimally small, we will focus on the intra-band paring components when we discuss the gap function.

\begin{figure}[htbp]
    \centering
    \includegraphics[width=0.36\textwidth]{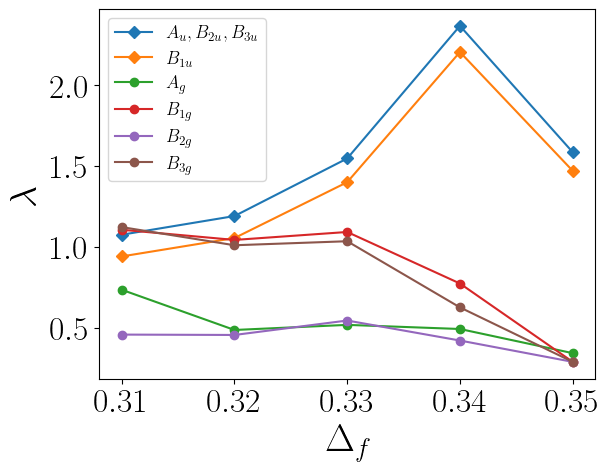}
    \caption{
    The $\Delta_f$ dependence of the superconducting eigenvalues for each IR in the absence of the SOC. 
    }
    \label{fig:lambda_Deltaf}
\end{figure}

First, we show the results without the SOC.
Figure~\ref{fig:lambda_Deltaf} illustrates the $\Delta_f$ dependence of superconducting eigenvalues, namely, the eigenvalues of the linearized Eliashberg equation for each IR of the $D_{2h}$ point group.
For a small $f$-electron level $\Delta_f$, the spin-singlet $B_{1g}$ and $B_{3g}$ states are stable, because the antiferromagnetic fluctuation works as an attractive force for spin-singlet superconductivity when the sign of gap function is opposite between the momenta connected by the nesting vector of Fermi surfaces~\cite{yanase2003Theory}. 
However, for large $\Delta_f$, the spin-triplet $A_u$, $B_{2u}$, and $B_{3u}$ states are stable, because the ferromagnetic fluctuation gradually becomes dominant and works as an attractive force for spin-triplet superconductivity.
These three spin-triplet superconducting states are degenerate due to spin $SU(2)$ symmetry in the absence of the SOC. 
In Fig.~\ref{fig:lambda_Deltaf}, we see that, even when the antiferromagnetic fluctuations are dominant ($\Delta_f = 0.31$ or $0.32$), the superconducting eigenvalues $\lambda$ for the spin-triplet superconducting states are comparable to or larger than those for the spin-singlet superconducting states. This is because the antiferromagnetic fluctuation also mediates an attractive force for spin-triplet Cooper pairs.
These results are consistent with our previous study of the single-sublattice periodic Anderson model~\cite{hakuno2024Magnetism}.

\begin{table}[h]
    \centering
    \caption{
    The leading order parameters of spin-triplet superconductivity in the absence of the SOC. 
    We show the magnitudes of the dominant gap function 
    $\sum_{\bm{k}}|d^{\mu}_{\nu}(\bm{k})|^2$ 
    integrated over the Brillouin zone. The $f$-electron level is set to $\Delta_f=0.32$.}
    \label{tab:without_SOC}
    \begin{tabular}{ccccccc}
        \hline
        IR & \multicolumn{6}{c}{Leading order parameter} \\ \hline
        $A_{u}$ & $d_z^0$ & $0.455$ & $d_z^x$ & $0.367$ & $d_z^y$ & $0.178$ \\
        $B_{2u}$ & $d_x^0$ & $0.455$ & $d_x^x$ & $0.367$ & $d_x^y$ & $0.178$ \\
        $B_{3u}$ & $d_y^0$ & $0.455$ & $d_y^x$ & $0.367$ & $d_y^y$ & $0.178$ \\
        $B_{1u}$ & $d_y^0$ & $0.562$ & $d_y^x$ & $0.411$ & $d_y^y$ & $0.027$ \\\hline
    \end{tabular}
\end{table}
We show the leading order parameters of spin-triplet superconductivity and the magnitudes of them in Table~\ref{tab:without_SOC}.
The result is not qualitatively changed by varying the parameters of the model.
The dominant component of the $d$-vector for each IR is $\Delta^{A_u}=d_z^0\sigma^zi\sigma^y\otimes\tau^0$, 
$\Delta^{B_{2u}}=d_x^0\sigma^x i\sigma^y\otimes\tau^0$, $\Delta^{B_{3u}}=d_y^0\sigma^y i\sigma^y\otimes\tau^0$, 
and $\Delta^{B_{1u}}=d_y^0\sigma^yi\sigma^y\otimes\tau^0$. 
The gap functions of the ${A_u}$, ${B_{2u}}$, and ${B_{3u}}$ states show the same momentum dependence because these states are degenerate and related by spin rotation. 
This is ensured by the spin $SU(2)$ symmetry; the spin can rotate independently of the crystal momentum in the absence of the SOC.

 \begin{figure}[htbp]
    \centering
    \includegraphics[width=0.48\textwidth]{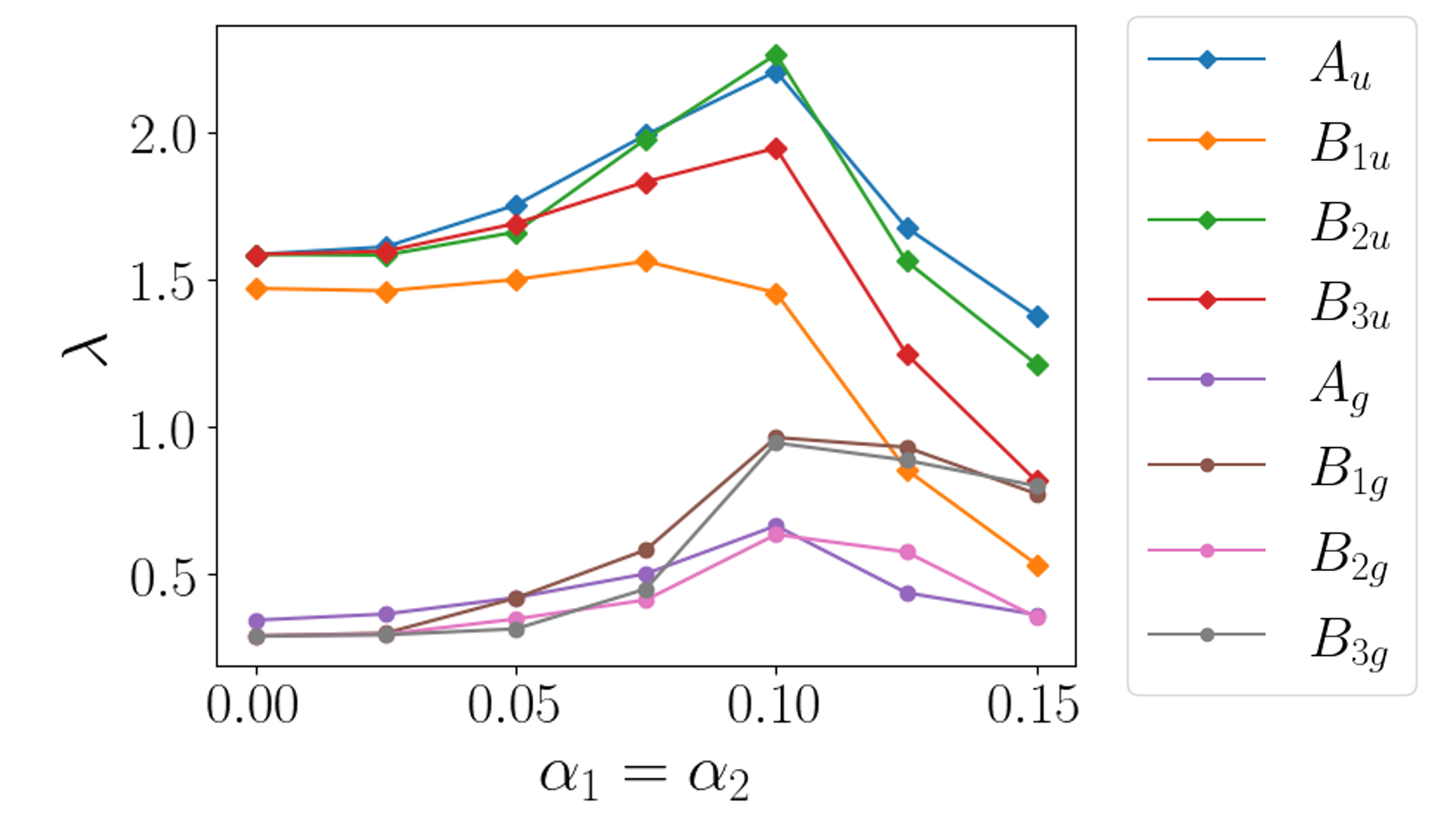}
    \caption{The SOC dependence of superconducting eigenvalues for each IR. We assume $\alpha_1 = \alpha_2$ and set $\Delta_f=0.35$.
    }
    \label{fig:ferromag_lambda}
\end{figure}

Next, we discuss superconductivity in the model with the SOC.
Figure~\ref{fig:ferromag_lambda} shows the superconducting eigenvalues as a function of the SOC. In this figure, we choose $\Delta_f =0.35$ and discuss the case in which ferromagnetic fluctuations are dominant in the absence of the SOC (see Fig.~\ref{fig:Delta_f_dep}).
The SOC lifts the degeneracy in terms of the spin rotational symmetry for the $A_u$, $B_{2u}$, and $B_{3u}$ states. We see that the $A_u$ state is stabilized by the SOC in a wide range of parameters, but the $B_{2u}$ state is comparable. The $A_u$ state is suggested by several experimental results, including the thermal transport measurement\cite{suetsugu2024Fully} and the NMR Knight shift measurement\cite{matsumura2023Large}.
The $B_{2u}$ state is also suggested by experiments such as the ultrasound measurement\cite{theuss2024Singlecomponent}. 
Spin-singlet superconductivity is not stable, although antiferromagnetic fluctuation becomes dominant in the large SOC region, as shown in Fig.~\ref{fig:susc_ferro}. This is probably because the spin fluctuation around $\bm{q}=\bm{0}$ works as a repulsive force for spin-singlet superconductivity and the magnetic anisotropy generally favors spin-triplet superconductivity~\cite{yanase2003Theory}.

\begin{table}[h]
    \centering
    \caption{The leading order parameters of superconductivity in the presence of the SOC. We show $\sum_{\bm{k}}|d^{\mu}_{\nu}(\bm{k})|^2$ calculated for $\alpha_1=\alpha_2=0.15$ and $\Delta_f=0.35$.}
    \label{tab:ferromag}
    \begin{tabular}{ccccccc}
        \hline
        IR & \multicolumn{6}{c}{Leading order parameter} \\ \hline
        $A_{u}$ & $d_z^y$ & $0.399$ & $d_x^0$ & $0.337$ & $d_x^x$ & $0.208$ \\
        $B_{2u}$ & $d_z^0$ & $0.392$ & $d_x^y$ & $0.289$ & $d_z^x$ & $0.286$ \\\hline
    \end{tabular}
\end{table}

The leading order parameters in the large SOC region are shown in Table~\ref{tab:ferromag}.
Although the $d_z^y\sigma^zi\sigma^y\otimes\tau^y$ component is largest for the $A_u$ representation, this component mainly gives the inter-band pairing, as is already mentioned. 
Therefore, the leading term appearing in the intra-band pairing is the $d_x^0\sigma^xi\sigma^y\otimes\tau^0$ component with the $d$-vector along the crystalline $a$ axis.
In contrast, the $d$-vector is predominantly directed toward the $c$ axis in the small SOC region. Thus, the dominant component of the $d$-vector in the $A_u$ state changes from the $\bm{d} \parallel c$ component to the $\bm{d} \parallel a$ component by increasing the SOC. 
The NMR measurements have reported a significant decrease in Knight shift along the $a$ axis below $T_{\rm c}$, indicating a large $a$-axis component of the $d$-vector\cite{matsumura2023Large}. Therefore, the $A_u$ state in the large SOC region is qualitatively consistent with the NMR experiment.
In contrast, in the $B_{2u}$ state, the dominant component of the $d$-vector changes from the $a$-axis component to the $c$-axis one as the SOC increases. Thus, the $B_{2u}$ state is consistent with the NMR experiment when we assume a small SOC.

 \begin{figure}[htbp]
    \centering
    \includegraphics[width=0.48\textwidth]{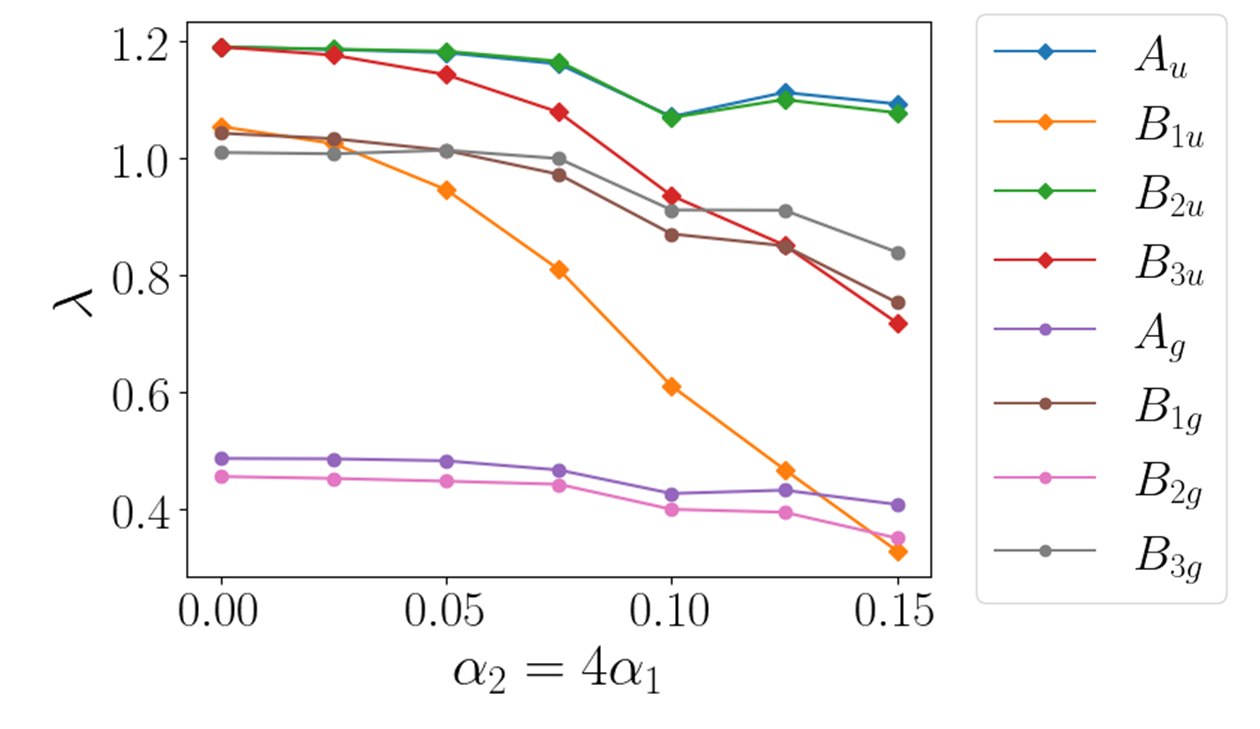}
    \caption{The SOC dependence of superconducting eigenvalues for each IR. We set $\Delta_f=0.32$ and $\alpha_2=4\alpha_1$.
    }
    \label{fig:antiferro_labmda}
\end{figure}

Here, we show the results of superconductivity when antiferromagnetic fluctuations are dominant.
Figure~\ref{fig:antiferro_labmda} shows the SOC dependence of superconducting eigenvalues for $\Delta_f=0.32$. 
While the $A_u$ and $B_{2u}$ states are almost degenerate and most stable up to $\alpha_2=4\alpha_1=0.15$, 
the $B_{3u}$ state is destabilized by the SOC. 
Spin-singlet superconductivity is relatively stable compared to the case with a ferromagnetic fluctuation (Fig.~\ref{fig:ferromag_lambda}), but it is also suppressed by the SOC.
This is consistent with the fact that spin-singlet superconductivity is strongly suppressed by magnetic anisotropy~\cite{sato2000Mechanism}. 
The leading order parameters obtained in the large SOC region are shown in Table~\ref{tab:antiferro}.
The $d_x^0\sigma^xi\sigma^y\otimes\tau^0$ component is dominant in the $A_u$ state.
The dominant component of the $d$-vector changes from the $c$-axis ($a$-axis) component to the $a$-axis ($c$-axis) component for the $A_u$ ($B_{2u}$) state, similarly to the results in Fig.~\ref{fig:ferromag_lambda}  for $\Delta_f=0.35$. Therefore, the above discussion for comparison to the NMR experiment also applies to this case.
Overall, the results are similar for both cases of ferromagnetic and antiferromagnetic fluctuations, except for the relative stability of spin-singlet superconductivity.

\begin{table}[h]
    \centering
    \caption{Leading order parameters of superconductivity for $4\alpha_1=\alpha_2=0.15$ and $\Delta_f=0.32$.}
    \label{tab:antiferro}
    \begin{tabular}{ccccccc}
        \hline
        IR & \multicolumn{6}{c}{Leading order parameter} \\ \hline
        $A_{u}$ & $d_x^0$ & $0.362$ & $d_z^y$ & $0.355$ & $d_x^x$ & $0.249$ \\
        $B_{2u}$ & $d_x^y$ & $0.362$ & $d_z^0$ & $0.358$ & $d_z^x$ & $0.249$ \\\hline
    \end{tabular}
\end{table}

\begin{figure}[htbp]
    \centering
    \includegraphics[width=0.50\textwidth]{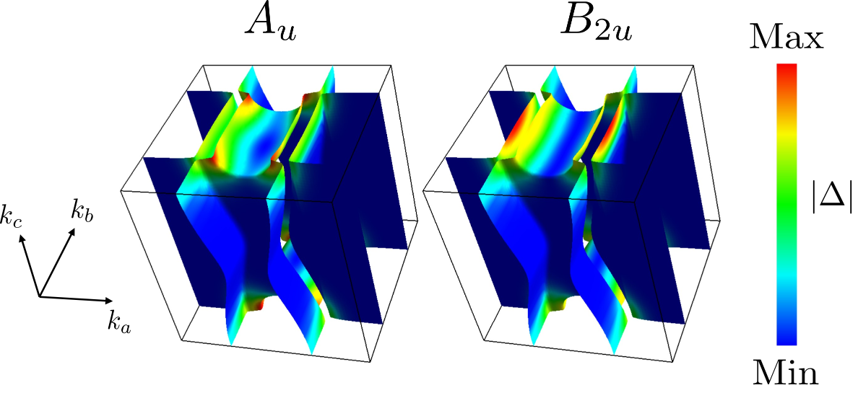}
    \caption{
    Superconducting gap function in the band representation for the $A_u$ (left) and $B_{2u}$ (right) states. The parameters are set as $\Delta_f=0.35$ and $\alpha_1=\alpha_2=0.15$.
    }
    \label{fig:gap_band}
\end{figure}

Figure~\ref{fig:gap_band} shows the absolute value of the gap function in the band representation for the $A_u$ and $B_{2u}$ superconducting states. The magnitude on the Fermi surfaces is highlighted by color.
We see that the superconducting gap has a pseudo-line node in both the $A_u$ and $B_{2u}$ states, which is not protected by symmetry. Though not clearly visible in the figure, the superconducting gap for the $B_{2u}$ state exhibits symmetry-protected point nodes at the intersection of the $b$-axis and the Fermi surface.

Finally, we discuss the effects of the SOC on superconductivity. 
The spin-triplet superconducting states that are stable in the strong SOC region can be understood by assuming that magnetic fluctuations along the spin direction perpendicular (parallel) to the $d$-vector induce an effectively attractive (repulsive) interaction in the spin-triplet channel. This assumption would be valid when the isotropic magnetic fluctuation stabilizes spin-triplet superconductivity in the absence of the SOC, as in the case of our model.
For the parameters in Figs.~\ref{fig:susc_ferro} and \ref{fig:susc_antiferro}, the spin susceptibility along the $b$ axis shows the maximum at $\bm{q}=(0,\pi,0)$ when the SOC is large.
Therefore, magnetic anisotropy favors the $\hat{x}$ and $\hat{z}$ components of the $d$-vector.
Antiferromagnetic fluctuations with $\bm{q}=(0,\pi,0)$ stabilize the orbital components $k_x$ and $k_z$ of the odd-parity Cooper pairs.
Combining the favorable spin and orbital components, we expect that the $A_u$ state with the $k_x \hat{x}$ and $k_z \hat{z}$ components in the gap function and the $B_{2u}$ state with $k_x \hat{z}$ and $k_z \hat{x}$ are favored, consistent with our numerical analysis of the linearized Eliashberg equation.
In contrast, the $B_{1u}$ and $B_{3u}$ states are suppressed because the gap function contains the $k_y$ orbital component or the $\hat{y}$ spin component. 
Therefore, the effects of the SOC are considered to manifest itself through magnetic anisotropy in this model.

\section{Summary and discussion}
In this paper, we constructed a spin-orbit-coupled periodic Anderson model with two sublattices in the unit cell. Taking into account the staggered SOC allowed in locally noncentrosymmetric structures, we calculated magnetic fluctuations and superconducting instability. Depending on the $f$-electron level, the antiferromagnetic fluctuation with $b$-axis anisotropy, the ferromagnetic fluctuation with $a$-axis anisotropy, and the simultaneous coexistence of the $a$-axis ferromagnetic fluctuation and the $b$- and $c$-axis antiferromagnetic fluctuations have been revealed.
The SOC lifts the degeneracy in the spin-triplet superconducting states protected by the $SU(2)$ spin rotation symmetry and stabilizes the superconducting states with the $A_u$ IR and $B_{2u}$ IR in a wide range of parameters. 
This anisotropy in the superconducting state can be understood on the basis of the momentum and directional dependence of magnetic fluctuations.
The direction of the superconducting $d$-vector for the $A_u$ ($B_{2u}$) state changes from $c$-axis to $a$-axis ($a$-axis to $c$-axis) by increasing the SOC, although the perpendicular components are small but finite.

While the thermal transport\cite{suetsugu2024Fully} and NMR\cite{matsumura2023Large}  measurements for a spin-triplet superconductor candidate UTe$_2$ suggest the superconducting state of $A_u$ IR, other experimental probes have indicated different symmetries: scanning tunneling microscope (STM) studies of surface states suggest the $B_{3u}$ state\cite{Gu_science2025,Wang2025}, ultrasound measurements~\cite{theuss2024Singlecomponent} of elastic moduli and specific heat measurements~\cite{Lee2025} point to the $B_{2u}$ state, and Josephson junction experiments have indicated the $B_{1u}$ state\cite{liObservationOddparitySuperconductivity2025}. 
Other constraints are also obtained by point contact spectroscopy~\cite{Yoon2024} and STM~\cite{yin2025yinyangvortexute2011}. 
Furthermore, based on the temperature and directional dependence of the magnetic penetration depth, the chiral $B_{3u}+iA_u$ state was proposed\cite{ishiharaChiralSuperconductivityUTe22023}.
Thus, the superconducting symmetry remains under debate. This is potentially due to the complex gap structure that non-trivially depends on the band and momentum. As illustrated in Fig.~\ref{fig:gap_band}, the band-resolved gap function calculated by many-body calculations generally differs from the simple basis functions listed in Table~\ref{tab:Irreps}~\cite{hakuno2024Magnetism,ishizuka2021Periodica,kreisel2022Spintriplet} and may require a quantitative comparison with experimental results. 
Recently, magnetic fluctuations were calculated based on a multi-orbital model derived from the DFT+$U$ method, which supported the antiferromagnetic fluctuation around ${\bm q} \sim (0,\pi,0)$~\cite{shimizu_private}. 
A future issue is the study of superconductivity in such a first-principles-based model.

In this work, the model parameters are chosen to be consistent with the DFT+$U$ calculation for UTe$_2$~\cite{ishizuka2019InsulatorMetal} and a phenomenologically renormalized model~\cite{liu2024densityfunctionaltheorybased}. 
Although our model does not explicitly take into account the effects of pressure and magnetic field, here we discuss some scenarios for a coherent understanding of multiple superconducting phases observed in UTe$_2$ under applied pressure and magnetic fields. 

The presence of at least two superconducting phases under pressure and zero magnetic field has been experimentally verified~\cite{braithwaiteMultiple2019,aokiMultiple2020}. Furthermore, AC magnetic susceptibility measurements\cite{kinjo2023Change,sakai2023Field} and specific heat measurements\cite{rosuel2023FieldInduced} have detected the transition line between the low-field and high-field superconducting phases in the $b$-axis magnetic field at ambient pressure.
Using ultra-pure single crystals with transition temperature $T_{\rm c}=2.1$K, specific heat measurements have shown that the superconducting state at high pressure and that in a high magnetic field are a thermodynamically equivalent phase~\cite{vasina2025connecting}. The NMR study in the normal state has shown that longitudinal magnetic fluctuations along the $b$ axis are enhanced~\cite{tokunaga2023Longitudinal} in the high magnetic field along the $b$ axis.
As demonstrated in our calculations, the magnetic fluctuation along the $b$ axis favors the $a$- and $c$-axis components of the $d$-vector, while suppressing the $b$-axis component. Thus, consistent with our results, a spin-triplet superconducting state of $B_{2u}$ or $A_u$ IR may be realized in the field- or pressure-induced superconducting phase. The NMR Knight shift studies in the superconducting state under pressure\cite{kinjo2023superconducting} or magnetic field\cite{kinjo2023Change} suggest the $d$-vector with substantial components along the $a$ and $c$ axes, consistent with the $B_{2u}$ and $A_u$ states obtained in this paper. 
Assuming this scenario, we next discuss the low-field superconducting phase at ambient pressure.
Under the magnetic field along the $b$ axis, the symmetry of the point group is reduced from $D_{2h}$ to $C^y_{2h}$.
Then, $A_u$ and $B_{2u}$ or $B_{1u}$ and $B_{3u}$ are reduced to the same IR, and they can not be distinguished by the symmetry~\cite{ishizuka2019InsulatorMetal}. 
Supposing the second-order phase transition lines dividing the high-field superconducting state from the low-field one, symmetry of the low-field phase is expected to be different from the high-field phase. In this case, the superconducting state at zero and low magnetic fields should be a $B_{1u}$ or $B_{3u}$ state, inconsistent with our results.
However, if a phase transition is first-order, the high- and low-field superconducting states can be either the $A_u$ or $B_{2u}$ state, consistent with our results.

\begin{acknowledgments}
We appreciate fruitful discussions with K. Nogaki, T. Kitamura, M. Shimizu, and J. Tei. This work was supported by JSPS
KAKENHI (Grant Numbers JP22H01181, JP22H04933,
JP23K17353, JP23K22452, JP24K21530, JP24H00007, JP25H01249).
We use FermiSurfer to describe the Fermi surfaces in this paper\cite{kawamura2019FermiSurfer}.

\end{acknowledgments}

\bibliography{ref}
 
\end{document}